\documentclass[fleqn,10pt]{wlscirep}
\usepackage[utf8]{inputenc}
\usepackage[T1]{fontenc}
\usepackage{graphicx}
\usepackage{subfigure}
\usepackage{textcomp,mathcomp}
\title{Silicon Optical Memory: Non-Volatile Optoelectronic Devices via Si-SiO$_2$  Hysteresis Effect}

\author[1]{Yuan Yuan}
\author[1,*]{Yiwei Peng}
\author[1]{Stanley Cheung}
\author[1]{Wayne V. Sorin}
\author[1]{Zhihong Huang}
\author[1]{Di Liang}
\author[1]{Marco Fiorentino}
\author[1]{Raymond G. Beausoleil}
\affil[1]{Hewlett Packard Labs, Hewlett Packard Enterprise, Milpitas, CA 95035, USA}

\affil[*]{yiwei.peng@hpe.com}

\begin{abstract}
Implementing on-chip non-volatile optical memories has long been an actively pursued goal, promising significant enhancements in the capability and energy efficiency of photonic integrated circuits. Here, a novel optical memory has been demonstrated exclusively using the semiconductor primary material, silicon. By manipulating the optoelectronic effect of this device, we introduce a hysteresis effect at the silicon-silicon oxide interface, which in turn demonstrates multi-level, non-volatile optical data storage with robust retention and endurance. This new silicon optical memory provides a distinctively simple and accessible route to realize optical data storage in standard silicon foundry processes.

\end{abstract}
\begin{document}

\flushbottom
\maketitle

\thispagestyle{empty}

\section*{Introduction}

Data storage has become a crucial requirement in modern photonic systems, encompassing applications like optical interconnects, optical computing, and photonic quantum circuits. Photonic data storage can significantly improve the performance of existing photonic systems by eliminating the latency associated with electronic memories, reducing the need for optoelectronic conversion, and minimizing static energy consumption. In optical computing, for instance, optical memories can eliminate the von-Neumann bottleneck by reducing the traffic between the optical processor and electrical memory \cite{shastri2021photonics}, and facilitate novel system architectures such as spiking neural networks \cite{zenke2015diverse} and in-memory optical computing \cite{rios2019memory}. However, achieving the storage of light poses a significant challenge due to the inherently weak interaction of photons. In response, many efforts have been made over the past decades to develop optical memories. Similar to electronic memories, optical memories are broadly classified into two main categories: volatile and non-volatile memories \cite{alexoudi2020optical}.
Volatile optical memories, that lose the stored data when the power is off, can be implemented through bistable optical devices \cite{liu2010ultra,kuramochi2014large}, optical ion excitation \cite{malacarne2007erbium}, and recirculating loop arrangements \cite{berrettini2011all}.
Non-volatile optical memories, on the other hand, can maintain the stored data without power. They can be achieved by altering the material properties interacting with light, including phase-change memories (PCMs)\cite{rios2015integrated, zhang2019broadband,zheng2018gst,fang2023non}, ferroelectric memories \cite{scott2007applications, mikolajick2020past,geler2021non}, micro-electro-mechanical systems (MEMS) \cite{honardoost2022low,han202132,quack2023integrated}, floating-gate memories \cite{barrios2006silicon,song2016integrated,zhu2023non,cheung2023ultra,cheung2023non}, and optical memristors \cite{tossoun2020memristor,cheung2022heterogeneous,youngblood2023integrated,cheung2023nonGFP}.
PCMs are characterized by thermal energy-dependent properties, transitioning between amorphous and crystalline states. This reversible transformation leads to significant changes in optical properties, affecting both the real and imaginary components of the complex refractive index. The representative PCM material is germanium-antimony-tellurium alloy (GST), which requires additional sputtering on photonic integrated circuits (PICs) \cite{zheng2018gst}.
Ferroelectric memories can toggle between two polarization states, which also require additional integration of ferroelectric materials, such as lead zirconium titanate (PZT), barium titanate (BTO), polyvinylidene flouride (PVDF), and hafnium oxide (HfO$_2$), on the Si chip.
MEMS devices employ various mechanisms, such as electrostatic, electrothermal, electromagnetic, and piezoelectric methods. Electrostatic actuation, in particular, allows low-power MEMS on a pure silicon (Si) platform. Extensive efforts have been dedicated to integrating MEMS into the standard Si photonics process, progressing from multi-layer MEMS to single-layer MEMS. However, the movable parts of MEMS still necessitate several additional processes for release \cite{han202132,quack2023integrated}. Additionally, the switching voltage of MEMS tends to be very high.
Optical floating-gate memories, akin to their electrical counterparts, involve a thin oxide dielectric layer and a floating gate, which are compatible with complementary-metal-oxide-semiconductor (CMOS) technology. The application of a bias voltage to the floating gate induces a high electric field, facilitating carrier injection into the oxide layer. Consequently, the reflective index of the device undergoes changes owing to the plasma dispersion effect. The current Si photonics foundries also require additional doping and Si/III-V bonding processes for optical floating-gate memories.
Optical memristors are also CMOS-compatible, which can be integrated on Si PICs through heterogeneous integration \cite{cheung2022heterogeneous}. But similar to floating-gate memories, an extra memristor layer is needed in the devices, such as HfO$_2$, zinc oxide (ZnO), titanium dioxide (TiO$_2$), tungsten oxide (WO$_x$), and tantalum pentoxide (Ta$_2$O$_5$) \cite{hu2023inkjet}. Besides that, the yield and reliability of memristors still face certain challenges.

Unlike the aforementioned technologies of non-volatile optical memories, here we demonstrate a novel optical memory exclusively utilizing the semiconductor primary material, Si. The Si optical memory is built on the standard Si photonics process, eliminating the need for additional materials, layers, or processes in the current Si photonics foundries. In the microelectronic community, bias temperature instability (BTI) has become a serious problem today as the size of metal-oxide-semiconductor field effect transistors (MOSFETs) shrinks. When the gate of a heated MOSFET is heavily biased, a strong electric field through the oxide layer results in the degradation of the MOSFET threshold voltage. Over the past decade, experimental evidence has confirmed that charge trapping is the cause of BTI \cite{stathis2006negative,hehenberger2011advanced,rzepa2014physical}. While BTI poses a significant challenge in microelectronics, the associated charge trapping phenomenon may hold promise for non-volatile memories. Here, we employ charge trapping at the Si-silicon oxide (SiO$_2$) interface to achieve Si optical memories. Different from MOSFETs, there is no gate or a vertical electric field on the oxide layer. Instead, traps at the Si-SiO$_2$ interface are charged by avalanche-multiplying photocurrent. 
This new mechanism allows for a hysteresis effect on the Si-SiO$_2$ interface, imparting non-volatile switching characteristics to the optical memory device. The program and erase states of the optical memory are implemented through a standard Si P-N junction, ensuring full compatibility with existing Si photonics devices. Consequently, the Si optical memories offer a groundbreaking approach to achieving photonic data storage. Leveraging the existing fabrication process, it presents a simple, cost-efficient, and high-yield pathway to integrate optical memories into Si PICs, directly applicable in areas such as optical interconnects, optical computing, and photonic quantum circuits.

\section*{Results}


\subsection*{Device structure}

\begin{figure}[ht]
\centering
\includegraphics[width=0.85\linewidth]{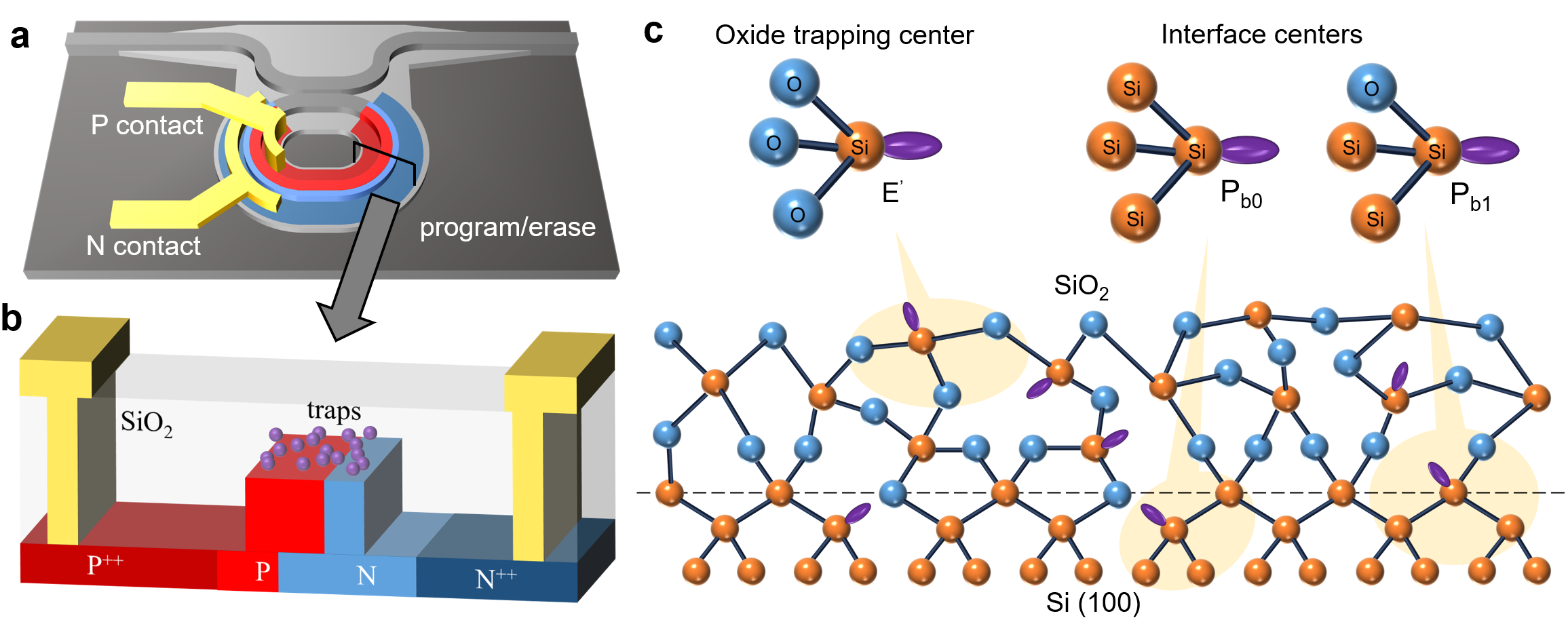}
\caption{\textbf{Si optical memory device structure.} \textbf{a}, Schematic diagram of the Si optical memory based on microring structure. \textbf{b}, Cross-sectional diagram of the doped microring waveguide. \textbf{c}, Configuration of Si-SiO$_2$ interface and corresponding traps.}
\label{fig.1}
\end{figure}

The schematic diagram of the non-volatile Si optical memory is presented in Fig. \ref{fig.1}a, the device is fabricated on a standard (100) orientation, 220 nm-thick Si-on-insulator (SOI) wafer. The optical memory incorporates an all-pass microring resonator (MRR) structure, with the MRR waveguide doped to form a Si P-N junction. The cross-section of the doped MRR waveguide is illustrated in Fig. \ref{fig.1}b, the waveguide core region is doped with P-type and N-type dopants to create a depletion interface inside the core region. A standard SiO$_2$ layer covers the entire Si all-pass MRR. The imperfect transition between crystalline Si and amorphous SiO$_2$ leads to some unbonded Si atoms, creating unpaired electrons that localize on the defect Si atoms and forming dangling bond traps. The configuration of Si-SiO$_2$ interface is depicted in Fig. \ref{fig.1}c, with potential traps including: 1) $E^{'}$ oxide trapping center, O$_3\equiv$Si$^\bullet$, which comprises of an unpaired electron localized on a Si atom while back-bonded three oxygen atoms; 2) $P_{b0}$ interface center, Si$_3\equiv$Si$^\bullet$, which involves an unpaired electron and back-bonded three Si atoms; and 3) $P_{b1}$ interface center, Si$_2$O$\equiv$Si$^\bullet$, characterized by an unpaired electron and back-bonded two Si atoms and one oxygen atom.
$E^{'}$ traps have levels near the middle of the SiO$_2$ bandgap, facilitating hole capture from the Si layer, thereby exhibiting donor-like traps \cite{entner2007modeling}. $P_{b0}$ and $P_{b1}$ trap centers only show donor-like states within the Si bandgap allowing for hole capture \cite{fussel1996defects}. These donor-like traps are electrically neutral when empty and positively charged when occupied by holes.
The occupation of the traps varies according to the stress conditions applied to the device. The charged traps in the Si-SiO$_2$ interface establish a localized electrical field that attracts a charge of the opposite polarity in the Si waveguide toward the interface. This space charge distribution relocates the doping profile of the Si P-N junction, thereby changing the effective refractive index of the waveguide through the Si plasma dispersion effect. As a result, by occupying/emptying the interface traps, the Si MRR waveguide exhibits different refractive indices and thus distinct resonant wavelengths in its optical spectrum. It is worth noting that the trap occupation ratio changes under specific stress conditions, therefore, once this device is programmed/erased, it will exhibit a multi-level non-volatile wavelength shift.


\subsection*{Program and erase characteristics}

\begin{figure}[ht]
\centering
\includegraphics[width=0.9\linewidth]{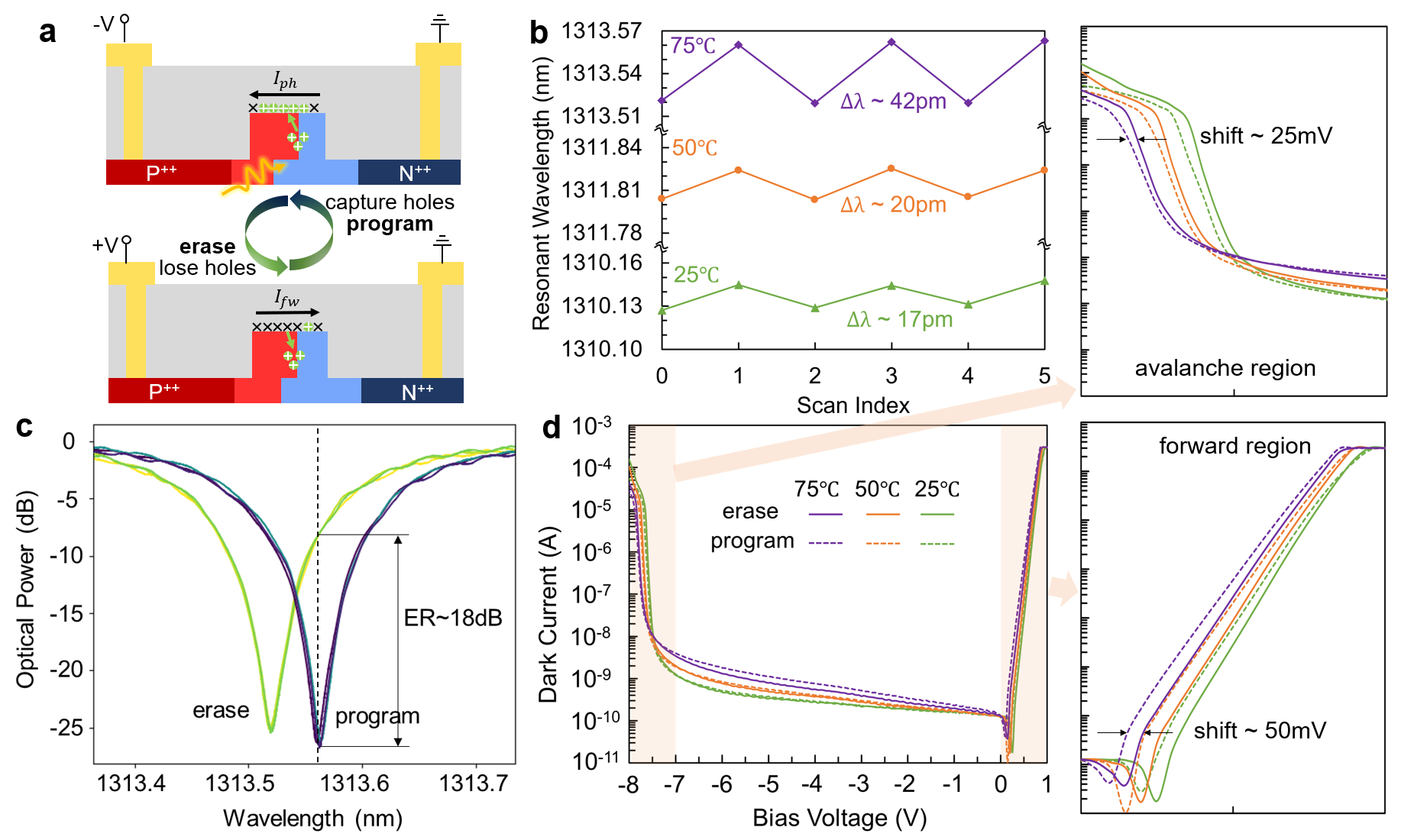}
\caption{\textbf{Program and erase characteristics of the Si optical memory.} \textbf{a}, Schematic diagram of the program and erase process of the Si optical memory. \textbf{b}, Measured resonant wavelength switches of the Si optical memory at 25 $\tccentigrade$ (green), 50 $\tccentigrade$ (orange), and 75 $\tccentigrade$ (purple). \textbf{c}, Measured optical spectrum of the Si optical memory at 75 $\tccentigrade$. \textbf{d}, Measured dark current of the Si optical memory at 25 $\tccentigrade$ (green), 50 $\tccentigrade$ (orange), and 75 $\tccentigrade$ (purple), along with expanded current curves in the avalanche region (program) and the forward region (erase).}
\label{fig.2}
\end{figure}

The program and erase conditions of this optical memory are depicted in Fig.\ref{fig.2}a. For programming, the Si P-N junction is reverse biased close to its avalanche breakdown voltage while an O-band laser is fed into the MRR to generate a considerable photocurrent $I_{ph}$ of $\sim$ 100 $\mu$A. This notable $I_{ph}$ arises from the combination of photon-assisted tunneling, resonant enhancement, and avalanche gain \cite{yuan2023mechanisms}. Under this condition, the Si-SiO$_2$ interface traps will gradually capture holes. The positively charged traps redistribute the doping profile of the waveguide, and the device exhibits the program state. The photocurrent is crucial during the program progress; otherwise, the resonant wavelength shift relying solely on high reverse bias voltage would be very small. This is because, to generate more free holes for trap capture, both photoillumination and impact ionization are required. Impact ionization itself, i.e., high reverse bias stress, cannot offer enough extra holes to significantly change the trap occupancy.
In contrast, only electrical stress is required during the erase progress. The Si P-N junction is forward biased to generate a forward current $I_{fw}$ of around hundreds $\mu$A. The large $I_{fw}$ can discharge the traps, therefore, reset the optical memory to the erase state. The detailed analysis is described in the Mechanisms and simulations section.

The switch in resonant wavelength of the optical memory are presented in Fig. \ref{fig.2}b, from bottom to top, measured at 25 $\tccentigrade$, 50 $\tccentigrade$, and 75 $\tccentigrade$, respectively. All data points are recorded with electrical power off, with even scan indices taken after erasing and odd scan indices after programming. The program state demonstrates a red-shifted resonant wavelength compared to the erase state across all three temperatures. Specifically, the resonant wavelength shift $\Delta\lambda$ is $\sim$ 17 pm at 25 $\tccentigrade$, $\sim$ 20 pm at 50 $\tccentigrade$, and $\sim$ 42 pm at 75 $\tccentigrade$. The difference between two states  $\Delta\lambda$ increases with temperature. This indicates that hole trapping is a temperature-dependent process. One possible explanation is that hole trapping is a non-radiative multi-phonon process (NMP) \cite{hehenberger2011advanced}, wherein the traps are required to be thermally excited before they can capture holes followed by structural relaxation, see Supplementary Information I.
Given that devices under 75 $\tccentigrade$ show the most significant memory shift, the subsequent results were measured at 75 $\tccentigrade$ unless otherwise specified. The corresponding transmission spectrum within three cycles of the optical memory at 75 $\tccentigrade$ is shown in Fig. \ref{fig.2}c. The three erase state curves and the three program state curves overlap well, respectively, indicating the optical memory device has stable and reproducible states. The red-shifted $\Delta\lambda$ of $\sim$ 42 pm results in an extinction ratio (ER) of $\sim$ 18 dB.

In addition to the non-volatile change in the optical domain, the device also exhibits a shift in the electrical domain. Figure \ref{fig.2}d shows the measured dark current versus bias voltage ranging from +1 V to -8 V, where the solid lines are the erase state and the dash lines represent the program state. First of all, the device in the program state has a higher dark current at reverse bias at all temperatures, possibly due to the charged traps leading to a higher surface leakage current. Besides that, a voltage shift phenomenon is observed in both the avalanche region and forward region. In the enlarged plot of the avalanche region, it is evident that the breakdown voltage of the programmed optical memory is approximately 25 mV higher than that of the device in the erase state. The charged traps repel dopants of the same polarity, causing an expansion of the depletion layer and consequently an increase in breakdown voltage. 
Conversely, the optical memory in the erase state exhibits a $\sim$ 50 mV higher forward bias voltage for the same current value compared to that in the program state. This voltage shift is attributed to the variation in the built-in voltage $V_{bi}$ of the Si P-N junction, which is expressed as
\begin{equation}
    V_{bi}=\frac{k_BT}{q}\ln\bigg(\frac{N_A N_D}{n_i^2}\bigg),
    \label{eq:1}
\end{equation}
where $k_B$ is the Boltzmann constant, $T$ is the temperature, $q$ is the elementary charge, $N_A$ is the acceptor concentration, $N_D$ is the donor concentration, and $n_i$ is the intrinsic carrier concentration. The charged traps in the program state repel the $N_A$ doping concentration, leading to a decrease in the $V_{bi}$. Consequently, the optical memory is an optoelectronic device that displays non-volatile memory effect in both optical and electrical domains.

\subsection*{Memory performance}

\begin{figure}[ht]
\centering
\includegraphics[width=0.98\linewidth]{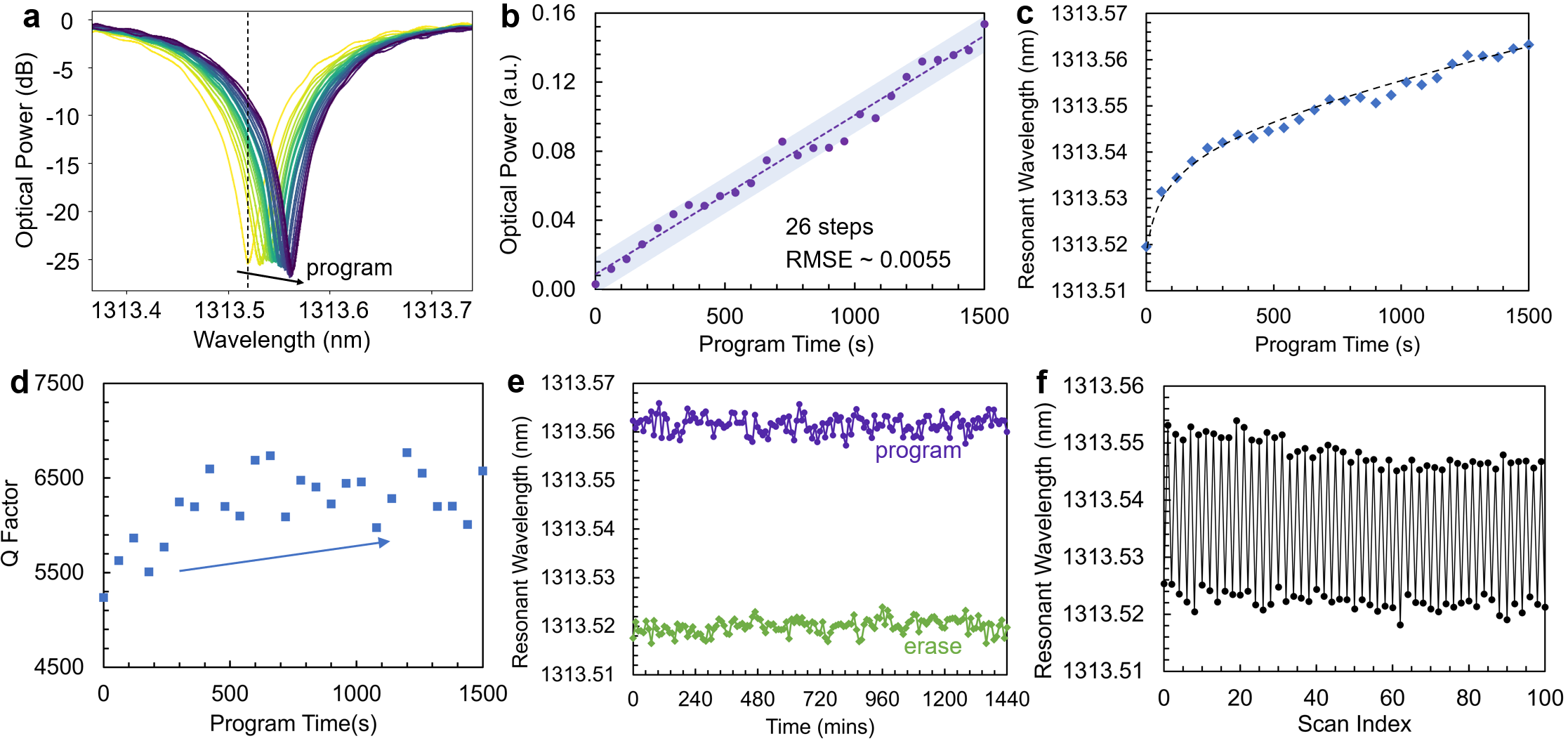}
\caption{\textbf{Performance of the Si optical memory.} \textbf{a}, Measured optical spectrum of the Si optical memory every 60 s of programming. \textbf{b}, Multiple optical power states versus program time of the Si optical memory at the wavelength of $\sim$ 1313.52 nm, as indicated by the black dash line in Fig. a. \textbf{c}, Corresponding resonant wavelengths versus program time. \textbf{d}, Corresponding quality factors (Q) of the microring Si optical memory from Lorentzian shape fitting. \textbf{e}, Retention measurements of non-volatile resonant wavelengths of the program and erase states within a 24-hour period. \textbf{f}, Endurance measurements of the Si optical memory with 50 cycles of program and erase.}
\label{fig.3}
\end{figure}

The Si P-N junction-based optical memory can be optoelectronically programmed and electronically erased, showcasing clear and repeatable switching capabilities. To quantify the performance of the memory, we conducted additional measurements. Figure \ref{fig.3}a illustrates the optical spectrum of the Si MRR recorded after every 60 s of programming. The optical spectrum gradually red-shifts with increased program time. Therefore, this optical memory exhibits multiple states depending on the program time. At the resonant wavelength of the erase state, i.e., the leftmost resonant wavelength of $\sim$ 1313.52 nm indicated by the black dash line in Fig. \ref{fig.3}a, the optical power levels of these multiple states are plotted in Fig. \ref{fig.3}b. The optical power transmission almost linearly increases with program time, this optical memory supports 26 states with optical power increases from $\sim$ 0.0 to 0.16. The root mean squared error (RMSE) of the 26 optical power states is around 0.0055. The multiple states of the optical memory are caused by the different occupations of the traps. The process of traps capturing holes can be described by a rate equation, thus the occupation of charged traps is a function of program time. Figure \ref{fig.3}c shows the corresponding resonant wavelengths from Fig. \ref{fig.3}a versus program time. As expected, the resonant wavelengths increase with program time similar to an exponential function $1-e^{-t}$. The red-shifted resonant wavelength indicates an increased refractive index of the waveguide. According to the well-known plasma dispersion effect of Si, changes in doping that result in an increase in refractive index also lead to a reduction in free carrier absorption loss. The corresponding Q factors are depicted in Fig. \ref{fig.3}d, aligning with the anticipated increasing trend versus program time. The fluctuation of the Q factors is attributed to uncertainties arising from the Lorentzian fitting of the optical spectrum.

One critical parameter for non-volatile memory is its retention capability. The resonant wavelengths of the optical memory were monitored every 10 minutes over a 24-hour period in both the program and erase states. As illustrated in Fig. \ref{fig.3}e, both program and erase states demonstrate stable resonant wavelengths for at least 24 hours. The program state has an average resonant wavelength of $\sim$ 1313.56 nm with a standard error of $\sim$ 1.49 pm, and the erase state has an average resonant wavelength of $\sim$ 1313.52 nm with a standard error of $\sim$ 1.24 pm. The remarkable stability over one day highlights the retention capability of the Si optical memory.
Cyclability of the memory is another important figure of merit. To evaluate this, the optical memory was subjected to a cycling test by alternately programming and erasing. The program operation is reverse biased near the breakdown voltage with a photocurrent $I_{ph}$ of $\sim$ 100 $\mu$A for 2500 s, while the erase operation is forward biased with hundreds $\mu$A current $I_{fw}$ for 600 s. Due to the relatively long cycle time, 50 cycles have been measured for this device. 
The measured resonant wavelengths after programming and erasing are plotted in Fig. \ref{fig.3}f, the optical memory shows consistent and repeatable switching behavior over 50 cycles, confirming its endurance capability.

\begin{figure}[ht]
\centering
\includegraphics[width=0.8\linewidth]{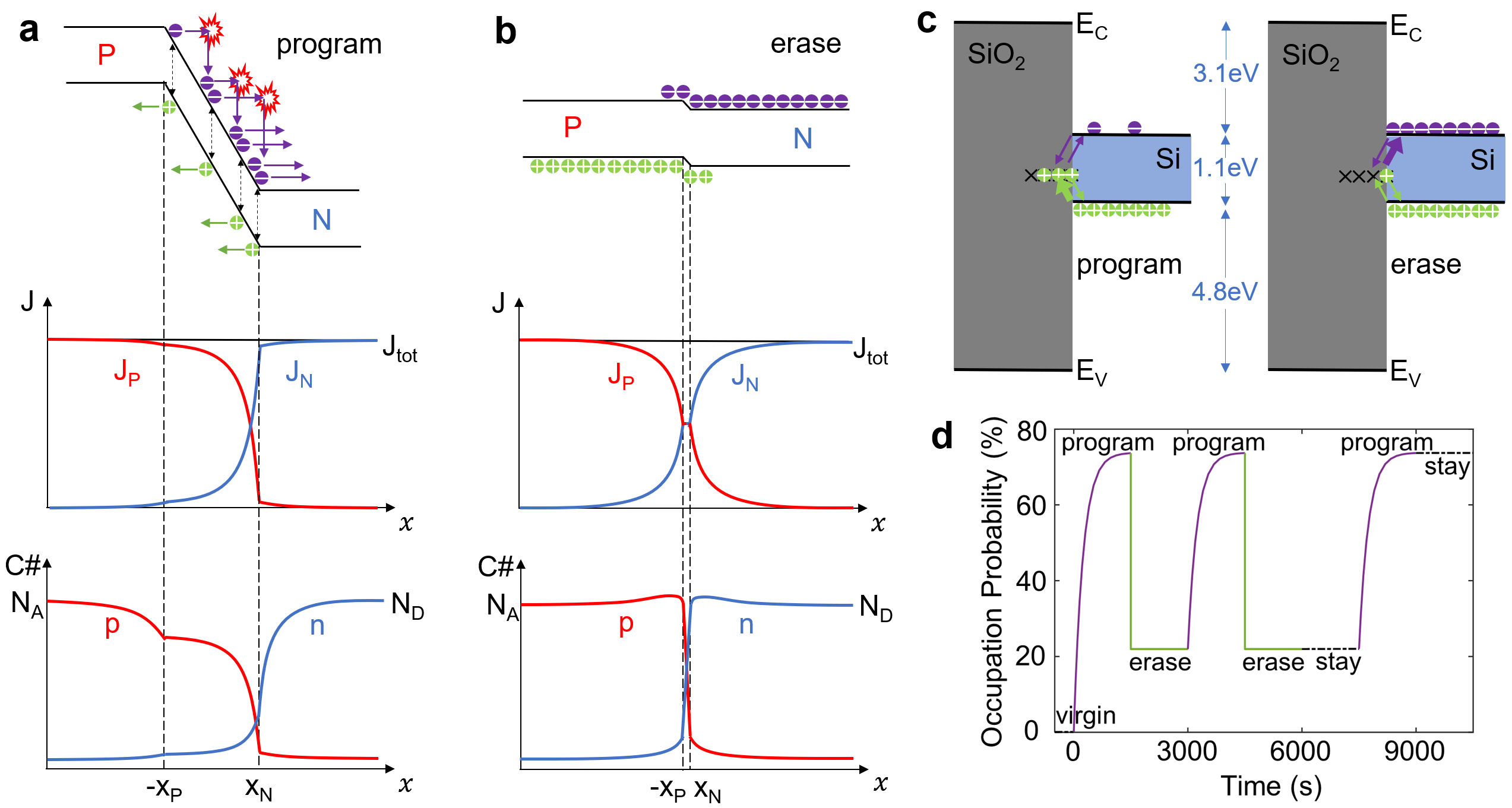}
\caption{\textbf{Program and erase mechanisms of the Si optical memory.} Schematic of the mechanism of the Si optical memory: band diagrams, current density distributions, and free carrier number distributions of the Si P-N junction during program (\textbf{a}) and erase (\textbf{b}) operations. \textbf{c}, Band diagrams of the Si-SiO$_2$ interface during program and erase operations. \textbf{d}, Simulated trap occupation probability of the Si optical memory at the virgin, program, erase, post-erase stay, and post-program stay states.}
\label{fig.4}
\end{figure}

\subsection*{Mechanisms and simulation}

The imperfect interface transition between (100) Si and amorphous SiO$_2$ introduces dangling bond traps, and these traps undergo filling and emptying during program and erase operations. The following charge switch of traps results in the hysteresis effect of the Si P-N junction-based waveguide. The transient behavior of donor-like traps is described by the rate equation
\begin{equation}
    \frac{dF_{tD}}{dt}=v_P\sigma_P\bigg[P(1-F_{tD})-F_{deg}F_{tD}n_i \exp\bigg(\frac{E_{tD}-E_i}{k_BT}\bigg)\bigg] - v_N\sigma_N\bigg[NF_{tD}-\frac{(1-F_{tD})n_i}{F_{deg}}\exp\bigg(\frac{E_i-E_{tD}}{k_BT}\bigg)\bigg],
    \label{eq:2}
\end{equation}
where $F_{tD}$ is the occupation probability of donor-like traps, $v_P$ and $v_N$ denote the thermal velocities for holes and electrons, $\sigma_P$ and $\sigma_N$ are the trap capture cross-sections for holes and electrons, $P$ and $N$ represent the concentrations of free holes and electrons, $F_{deg}$ is the degeneracy factor, $n_i$ is the intrinsic carrier concentration, $E_{tD}$ and $E_i$ are the donor-like traps and intrinsic energy levels \cite{silvaco2010}. There are four distinct terms on the right side of Eq. \ref{eq:2}: 1) a charge rate by capturing valence band holes, 2) a discharge rate to the valence band, 3) a discharge rate through conduction band electrons, and 4) a charge rate due to the emission of electrons to the conduction band. These charge/discharge rates linearly depend on the trap cross-sections $\sigma_P$ and $\sigma_N$, which describe the interaction between carriers and traps. Therefore, $\sigma_P$ and $\sigma_N$ are functions of trap location distance $d$ into the SiO$_2$,
\begin{eqnarray}
    \sigma_P(d)=\sigma_{P0}\exp(-\kappa_h d), \nonumber \\
    \sigma_N(d)=\sigma_{N0}\exp(-\kappa_e d),
    \label{eq:3}
\end{eqnarray}
where the constants $\sigma_{P0}$ and $\sigma_{N0}$ are about 10$^{-15}$ cm$^2$ and 10$^{-17}$ cm$^2$, respectively\cite{baba1986mechanism,hofmann1985hot}. $\kappa_h$ and $\kappa_e$ are the evanescent wavectors for electrons and holes,
\begin{eqnarray}
    \kappa_h^2=\frac{2m_h^*(E_{tD}-E_V)}{\hbar^2}, \nonumber \\
    \kappa_e^2=\frac{2m_e^*(E_C-E_{tD})}{\hbar^2},
    \label{eq:4}
\end{eqnarray}
where $m_h^*$ and $m_e^*$ are the effective masses of holes and electrons, $E_V$ and $E_C$ stand for the conduction band edge and valence band edge. In the subsequent simulations, it is assumed that the trap spatial density is uniform up to a certain depth and zero above that, and the traps are considered to be monoenergetic. The band diagram of the Si P-N junction under program operation is illustrated in Fig. \ref{fig.4}a. At a high reverse bias close to the breakdown voltage, avalanche gain occurs inside the depletion region. The photon-generated carriers accelerate within the depletion region due to the high electric field, leading to the impact ionization events. Due to the significant difference between the ionization coefficients of holes ($\beta$) and electrons ($\alpha$) in Si, impact ionization predominantly happens in electrons \cite{tan2000avalanche, kang2009monolithic}. The current density contributed by electrons, $J_N$, exhibits an exponential-like growth from left to right due to the increasing number of impact ionization-generated electrons. As the current density is uniform across the entire junction, the complementary hole current density, $J_P$, is represented by the red curve in the middle chart of Fig. \ref{fig.4}a. The average $J_P>J_N$ inside the depletion region because of the electron-perferred impact ionization. The electron and hole number distribution can then be extracted from the current density by $N=J_N/qV_{dN}$ and $P=J_P/qV_{dP}$, where $V_{dN}$ and $V_{dP}$ are the drift velocities of the electrons and holes. As $J_P>J_N$ and $V_{dN}\gg V_{dP}$, the hole concentration $P$ is much higher than the electron concentration $N$ inside the depletion region. Therefore, the first term on the right side of Eq. \ref{eq:2}, $V_P\sigma_PP(1-F_{tD})$, dominates the transient behavior. The Si-SiO$_2$ traps get charged by capturing holes.
Conversely, the Si P-N junction is forward-biased to erase the memory. The band diagram and current density distribution of the erase-operated junction are shown in Fig. \ref{fig.4}b. The $J_N$ and $J_P$ are almost at the same level due to the similar acceptor and donor doping concentrations, $N_A$ and $N_P$, inside the Si junction. The electron concentration at $-x_P$ is $N(-x_P)=\frac{n_i^2}{N_A}e^{qV/k_BT}$ and the hole concentration at $x_N$ is $P(x_N)=\frac{n_i^2}{N_D}e^{qV/k_BT}$, indicating that the $P$ and $N$ have similar concentrations. Therefore, the discharge term contributed by electrons is not negligible, the rate equation will lead to a new balance state with a smaller $F_{tD}$.

The band diagrams of the Si-SiO$_2$ interface during program and erase operations are shown in Fig. \ref{fig.4}c. The traps are charged due to the much higher $P$ concentration during the program operation, and they are discharged as the $P\sim N$ during the erase operation. The simulated occupation probability of traps, $F_{tD}$, is plotted in Fig. \ref{fig.4}d. In the virgin state, the device has not been probed at all, the traps are empty. During the program operation, the occupation probability gradually increases with time, and the saturation time is on the order of 10$^3$ s. On the contrary, the erase operation discharges the traps, where the erase speed is faster than the program due to the higher free carrier concentrations. The $F_{tD}$ quickly reduced from $\sim$ 74\% to $\sim$ 22\%, and then reaches a steady state. To show the non-volatility of this Si optical memory, we also simulate the steady state after the erase and program operations, where $P$ and $N$ values are zero as there is no bias and current through the device. The occupation probability of traps remains constant without voltage.

\begin{figure}[ht]
\centering
\includegraphics[width=1\linewidth]{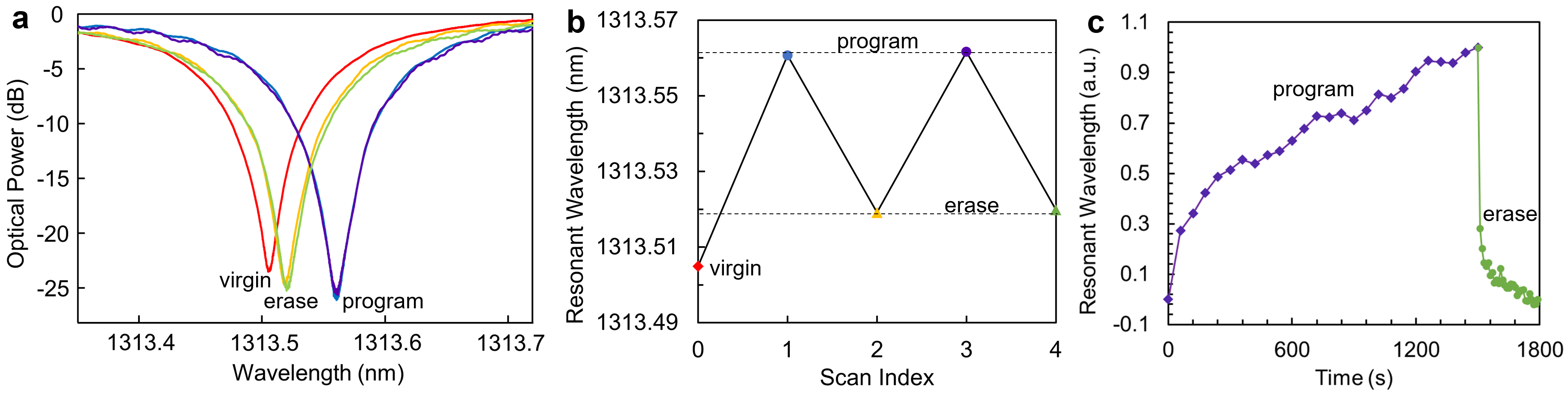}
\caption{\textbf{Mechanism validation measurements of the Si optical memory.} \textbf{a}, Measured optical spectrum of the virgin, erase, and program states of the Si optical memory. \textbf{b}, Corresponding resonant wavelength of the Si optical memory. \textbf{c}, Resonant wavelength versus time during program (purple) and erase (green) operations.}
\label{fig.5}
\end{figure}

In order to validate the model of the Si optical memory, a new device on the same chip has been measured to demonstrate the virgin, program, and erase states. Figure \ref{fig.5}a shows the measured optical spectrum of the new device, where the virgin state exhibits the shortest resonant wavelength, the program state shows the longest resonant wavelength, and the erase state has a slightly red-shifted resonant wavelength compared to the virgin state. The measured results align well with our simulated occupation probability in Fig. \ref{fig.4}d, the erase operation cannot fully empty the charged traps. The corresponding resonant wavelength is plotted in Fig. \ref{fig.5}b, the optical memory can program/erase several times, allowing a repeatable non-volatile wavelength shift of more than 40 pm. The program and erase speeds are also measured to further verify our model. As depicted in Fig. \ref{fig.5}c, the resonant wavelength versus operation time is measured every 60 s during the program (purple dots) and every 10 s during the erase (green dots) operations. The program operation takes about 1500 s to fully shift the Si optical memory, while the erase operation only requires 300 s to set the memory resonant wavelength back. The speed measurement also agrees well with the simulated results of our model.
It is noteworthy that $F_{tD}$ is a normalized probability, the resonant wavelength shift is also influenced by the trap density at the Si-SiO$_2$ interface. While the two Si optical memories reported here demonstrate a comparable non-volatile resonant wavelength shift of $\sim$ 42 pm, the hero device on this chip enables a shift of $\sim$ 71 pm, corresponding to an ultra-high ER of $\sim$ 26 dB. Detailed spectrum and resonant wavelengths are provided in Supplementary Information II.

\begin{figure}[ht]
\centering
\includegraphics[width=0.72\linewidth]{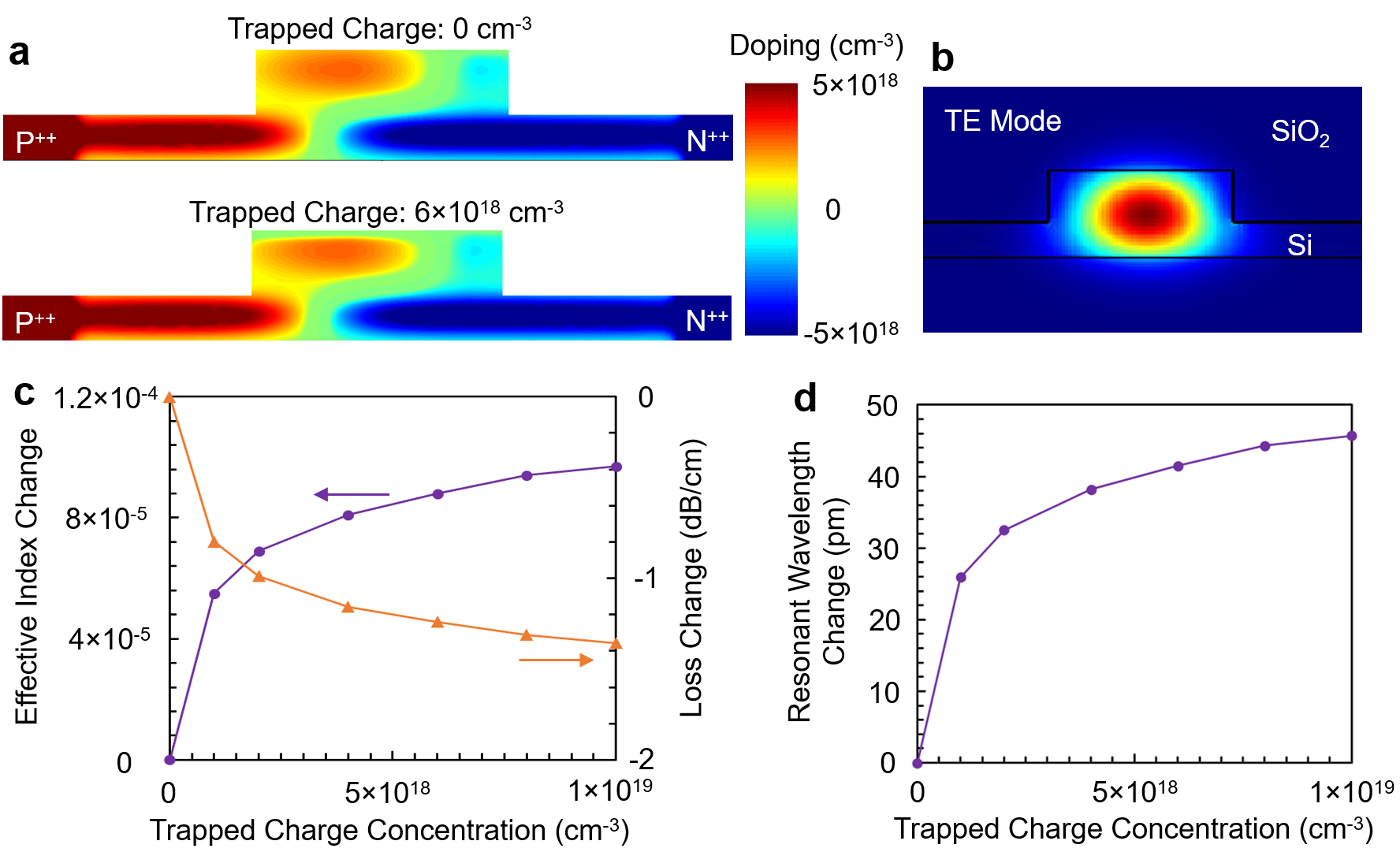}
\caption{\textbf{Simulation results of the Si optical memory.} \textbf{a}, Simulated doping concentration distributions of the Si P-N junction with 0 cm$^{-3}$ and 6$\times$10$^{18}$ cm$^{-3}$ charged traps. \textbf{b}, Transverse electric (TE) mode energy distribution inside the Si P-N junction waveguide. \textbf{c}, Simulated effective index change (purple) and loss change (orange) of the Si P-N junction waveguide versus trapped charge concentration. \textbf{d}, Corresponding resonant wavelength change of the Si microring resonator memory versus trapped charge concentration.}
\label{fig.6}
\end{figure}

The occupied traps at the Si-SiO$_2$ interface place a fixed charge on the surface of the Si waveguide. This charge induces a doping redistribution within the Si P-N junction, thereby altering the refractive index of the waveguide. In Fig. \ref{fig.6}a, the simulated doping concentration profiles of the Si P-N junction are compared without trapped charge and with 6$\times$10$^{18}$ cm$^{-3}$ charged traps. Given that the occupied traps carry positive charges, the P-type dopants are repelled from the surface, while the N-type dopants are attracted close to the surface. The corresponding transverse electric (TE) optical mode of the Si P-N junction-based waveguide can then be simulated for different doping profiles, as depicted in Fig. \ref{fig.6}b. The traced effective refractive index change versus trapped charge concentrations is plotted in Fig. \ref{fig.6}c. With 1$\times$10$^{19}$ cm$^{-3}$ charged traps, the effective index increases by approximately 9.7$\times$10$^{-5}$. At the same time, the free carrier absorption loss of the Si waveguide is reduced. As the orange curve shows, the loss reduces by $\sim$ 1.4 dB/cm. This reduction trend in loss agrees with the increased Q factor during the program operation. The calculated resonant wavelength change, based on the simulated effective index change, is presented in Fig. \ref{fig.6}d, where the Si MRR memory could shift more than 45 pm with 1$\times$10$^{19}$ cm$^{-3}$ charged traps.

\section*{Discussion}

In this work, we demonstrate a brand-new non-volatile optical memory based on standard Si photonic devices. Diverging from existing technologies, this Si optical memory leverages disparate current densities of holes and electrons from avalanche-multiplying photocurrent to generate hysteresis trapping charge at Si-SiO$_2$ interface. With no alterations to the fabrication processes, this Si optical memory maintains full compatibility with Si photonics foundries, making it seamlessly integrated into standard process design kit (PDK) and directly applicable in Si PICs. Furthermore, this Si optical memory is a generic device that is not contingent on the specifications of a particular Si photonics foundry. While the Si optical memories reported in the main text were produced at Advanced Micro Foundry (AMF), we have successfully replicated a similar non-volatile memory phenomenon in a Si P-N junction-based MRR fabricated at CEA-Leti. The detailed results for the Leti-fabricated optical memory are provided in Supplementary Information III.

At 75 $\tccentigrade$, the showcased Si optical memory exhibits a non-volatile shift of approximately 42 pm in its resonant wavelength, yielding a commendable ER of $\sim$ 18 dB. The extent of non-volatile shift depends on the interface trap density, and the leading device on the wafer can attain an impressive resonant wavelength shift of $\sim$ 71 pm with an ultra-high ER of $\sim$ 26 dB. Consequently, enhancing the Si optical memory shift is achievable through the engineering of the interface trap density. This Si optical memory also facilitates multiple states, with 26 distinct states achieved by varying the program time. The retention performance of the Si optical memory is commendable, as both the program and erase states maintain a highly stable resonant wavelength for at least 24 hours. Furthermore, the cyclability of the Si optical memory is validated through 50 program-erase cycles, affirming the device's endurance and demonstrating consistent and repeatable switching behavior. 
A charge trapping model has been developed for this optical memory. The rate equation of the occupation probability of donor-like traps highly depends on the concentrations of free electrons and holes. Given that electrons are much easier to impact ionization than holes in Si, and the drift velocity of electrons is significantly faster than that of holes, the free holes concentration is way larger than that of free electrons within the device during the program operation. The surplus holes thus effectively fill the donor-like traps, establishing a non-volatile charge difference at the Si-SiO$_2$ interface. This fixed charge offset, in turn, alters the doping profile inside the Si junction, interacting with the optical mode within the waveguide. The simulated results agree well with the measurements, offering an elucidation for the observed behavior of this innovative optical memory. In summary, a novel optical memory is demonstrated on Si photonics platform with zero change of standard processes, paving a simpler and most cost-efficient pathway to enhance the performance of existing PICs by eliminating the electronic memory latency, reducing times for optoelectronic conversion, and minimizing static energy consumption. 

\section*{Methods}
\subsection*{Fabrication}
The Si optical memories were fabricated at Advanced Micro Foundry (AMF), Singapore. The chips are based on industry-standard 220 nm-thick SOI wafers. The Z-shape junction was formed from Phosphorous (for N-type) and Boron (for P-type) implants.

\subsection*{Measurements}
The bias voltage was supplied by a Keithley 2400 source meter. An O-band tunable laser Santec TSL-550 was used as the light source for photocurrent and optical spectrum. All spectrum and resonant wavelengths were measured without bias voltage for non-volatile performance. The optical spectrum of the multiple states was measured without bias voltage after every 60 s of programming. The retention results were measured after a one-time program/erase operation, and then the spectrum was recorded without bias voltage every 10 minutes over 24 hours. The endurance results for 50 cycles are measured with an automatic setup that uses the same stress conditions for all program/erase operations.

\subsection*{Simulations}
The rate equation of donor-like traps was solved by the solve nonstiff differential equations - low order method (ode23) in Matlab. The donor-like traps energy level, $E_{tD}$, was chosen to be 0.5 eV. The free carrier concentrations $P$ and $N$ were calculated from the current densities during program and erase operations. The doping profiles of the Si P-N junction under different trapped charges, shown in Fig. \ref{fig.6}a, were simulated by Lumerical Charge. The simulated doping profiles were then imported into Lumerical Mode to simulate the corresponding TE mode in Fig. \ref{fig.6}b and extract effective refractive index change and loss change of the Si waveguide.

\section*{Data availability}
The data that support the findings of this study are available from the corresponding author upon request.

\bibliography{main}

\begin{thebibliography}{10}
\urlstyle{rm}
\expandafter\ifx\csname url\endcsname\relax
  \def\url#1{\texttt{#1}}\fi
\expandafter\ifx\csname urlprefix\endcsname\relax\def\urlprefix{URL }\fi
\expandafter\ifx\csname doiprefix\endcsname\relax\def\doiprefix{DOI: }\fi
\providecommand{\bibinfo}[2]{#2}
\providecommand{\eprint}[2][]{\url{#2}}

\bibitem{shastri2021photonics}
\bibinfo{author}{Shastri, B.~J.} \emph{et~al.}
\newblock \bibinfo{journal}{\bibinfo{title}{Photonics for artificial intelligence and neuromorphic computing}}.
\newblock {\emph{\JournalTitle{Nature Photonics}}} \textbf{\bibinfo{volume}{15}}, \bibinfo{pages}{102--114} (\bibinfo{year}{2021}).

\bibitem{zenke2015diverse}
\bibinfo{author}{Zenke, F.}, \bibinfo{author}{Agnes, E.~J.} \& \bibinfo{author}{Gerstner, W.}
\newblock \bibinfo{journal}{\bibinfo{title}{Diverse synaptic plasticity mechanisms orchestrated to form and retrieve memories in spiking neural networks}}.
\newblock {\emph{\JournalTitle{Nature communications}}} \textbf{\bibinfo{volume}{6}}, \bibinfo{pages}{6922} (\bibinfo{year}{2015}).

\bibitem{rios2019memory}
\bibinfo{author}{R{\'\i}os, C.} \emph{et~al.}
\newblock \bibinfo{journal}{\bibinfo{title}{In-memory computing on a photonic platform}}.
\newblock {\emph{\JournalTitle{Science advances}}} \textbf{\bibinfo{volume}{5}}, \bibinfo{pages}{eaau5759} (\bibinfo{year}{2019}).

\bibitem{alexoudi2020optical}
\bibinfo{author}{Alexoudi, T.}, \bibinfo{author}{Kanellos, G.~T.} \& \bibinfo{author}{Pleros, N.}
\newblock \bibinfo{journal}{\bibinfo{title}{Optical ram and integrated optical memories: a survey}}.
\newblock {\emph{\JournalTitle{Light: Science \& Applications}}} \textbf{\bibinfo{volume}{9}}, \bibinfo{pages}{91} (\bibinfo{year}{2020}).

\bibitem{liu2010ultra}
\bibinfo{author}{Liu, L.} \emph{et~al.}
\newblock \bibinfo{journal}{\bibinfo{title}{An ultra-small, low-power, all-optical flip-flop memory on a silicon chip}}.
\newblock {\emph{\JournalTitle{Nature Photonics}}} \textbf{\bibinfo{volume}{4}}, \bibinfo{pages}{182--187} (\bibinfo{year}{2010}).

\bibitem{kuramochi2014large}
\bibinfo{author}{Kuramochi, E.} \emph{et~al.}
\newblock \bibinfo{journal}{\bibinfo{title}{Large-scale integration of wavelength-addressable all-optical memories on a photonic crystal chip}}.
\newblock {\emph{\JournalTitle{Nature Photonics}}} \textbf{\bibinfo{volume}{8}}, \bibinfo{pages}{474--481} (\bibinfo{year}{2014}).

\bibitem{malacarne2007erbium}
\bibinfo{author}{Malacarne, A.}, \bibinfo{author}{Bogoni, A.} \& \bibinfo{author}{Poti, L.}
\newblock \bibinfo{journal}{\bibinfo{title}{Erbium--ytterbium-doped fiber-based optical flip-flop}}.
\newblock {\emph{\JournalTitle{IEEE Photonics Technology Letters}}} \textbf{\bibinfo{volume}{19}}, \bibinfo{pages}{904--906} (\bibinfo{year}{2007}).

\bibitem{berrettini2011all}
\bibinfo{author}{Berrettini, G.} \emph{et~al.}
\newblock \bibinfo{journal}{\bibinfo{title}{All-optical digital circuits exploiting soa-based loop memories}}.
\newblock {\emph{\JournalTitle{IEEE Journal of Selected Topics in Quantum Electronics}}} \textbf{\bibinfo{volume}{18}}, \bibinfo{pages}{847--858} (\bibinfo{year}{2011}).

\bibitem{rios2015integrated}
\bibinfo{author}{R{\'\i}os, C.} \emph{et~al.}
\newblock \bibinfo{journal}{\bibinfo{title}{Integrated all-photonic non-volatile multi-level memory}}.
\newblock {\emph{\JournalTitle{Nature photonics}}} \textbf{\bibinfo{volume}{9}}, \bibinfo{pages}{725--732} (\bibinfo{year}{2015}).

\bibitem{zhang2019broadband}
\bibinfo{author}{Zhang, Y.} \emph{et~al.}
\newblock \bibinfo{journal}{\bibinfo{title}{Broadband transparent optical phase change materials for high-performance nonvolatile photonics}}.
\newblock {\emph{\JournalTitle{Nature communications}}} \textbf{\bibinfo{volume}{10}}, \bibinfo{pages}{4279} (\bibinfo{year}{2019}).

\bibitem{zheng2018gst}
\bibinfo{author}{Zheng, J.} \emph{et~al.}
\newblock \bibinfo{journal}{\bibinfo{title}{Gst-on-silicon hybrid nanophotonic integrated circuits: a non-volatile quasi-continuously reprogrammable platform}}.
\newblock {\emph{\JournalTitle{Optical Materials Express}}} \textbf{\bibinfo{volume}{8}}, \bibinfo{pages}{1551--1561} (\bibinfo{year}{2018}).

\bibitem{fang2023non}
\bibinfo{author}{Fang, Z.} \emph{et~al.}
\newblock \bibinfo{journal}{\bibinfo{title}{Non-volatile materials for programmable photonics}}.
\newblock {\emph{\JournalTitle{APL Materials}}} \textbf{\bibinfo{volume}{11}} (\bibinfo{year}{2023}).

\bibitem{scott2007applications}
\bibinfo{author}{Scott, J.}
\newblock \bibinfo{journal}{\bibinfo{title}{Applications of modern ferroelectrics}}.
\newblock {\emph{\JournalTitle{science}}} \textbf{\bibinfo{volume}{315}}, \bibinfo{pages}{954--959} (\bibinfo{year}{2007}).

\bibitem{mikolajick2020past}
\bibinfo{author}{Mikolajick, T.}, \bibinfo{author}{Schroeder, U.} \& \bibinfo{author}{Slesazeck, S.}
\newblock \bibinfo{journal}{\bibinfo{title}{The past, the present, and the future of ferroelectric memories}}.
\newblock {\emph{\JournalTitle{IEEE Transactions on Electron Devices}}} \textbf{\bibinfo{volume}{67}}, \bibinfo{pages}{1434--1443} (\bibinfo{year}{2020}).

\bibitem{geler2021non}
\bibinfo{author}{Geler-Kremer, J.} \emph{et~al.}
\newblock \bibinfo{title}{A non-volatile optical memory in silicon photonics}.
\newblock In \emph{\bibinfo{booktitle}{2021 Optical Fiber Communications Conference and Exhibition (OFC)}}, \bibinfo{pages}{1--3} (\bibinfo{organization}{IEEE}, \bibinfo{year}{2021}).

\bibitem{honardoost2022low}
\bibinfo{author}{Honardoost, A.}, \bibinfo{author}{Henriksson, J.}, \bibinfo{author}{Kwon, K.}, \bibinfo{author}{Luo, J.} \& \bibinfo{author}{Wu, M.~C.}
\newblock \bibinfo{title}{Low-loss wafer-bonded silicon photonic mems switches}.
\newblock In \emph{\bibinfo{booktitle}{2022 Optical Fiber Communications Conference and Exhibition (OFC)}}, \bibinfo{pages}{1--3} (\bibinfo{organization}{IEEE}, \bibinfo{year}{2022}).

\bibitem{han202132}
\bibinfo{author}{Han, S.} \emph{et~al.}
\newblock \bibinfo{journal}{\bibinfo{title}{32$\times$ 32 silicon photonic mems switch with gap-adjustable directional couplers fabricated in commercial cmos foundry}}.
\newblock {\emph{\JournalTitle{journal of optical microsystems}}} \textbf{\bibinfo{volume}{1}}, \bibinfo{pages}{024003--024003} (\bibinfo{year}{2021}).

\bibitem{quack2023integrated}
\bibinfo{author}{Quack, N.} \emph{et~al.}
\newblock \bibinfo{journal}{\bibinfo{title}{Integrated silicon photonic mems}}.
\newblock {\emph{\JournalTitle{Microsystems \& Nanoengineering}}} \textbf{\bibinfo{volume}{9}}, \bibinfo{pages}{27} (\bibinfo{year}{2023}).

\bibitem{barrios2006silicon}
\bibinfo{author}{Barrios, C.~A.} \& \bibinfo{author}{Lipson, M.}
\newblock \bibinfo{journal}{\bibinfo{title}{Silicon photonic read-only memory}}.
\newblock {\emph{\JournalTitle{Journal of lightwave technology}}} \textbf{\bibinfo{volume}{24}}, \bibinfo{pages}{2898} (\bibinfo{year}{2006}).

\bibitem{song2016integrated}
\bibinfo{author}{Song, J.-F.} \emph{et~al.}
\newblock \bibinfo{journal}{\bibinfo{title}{Integrated photonics with programmable non-volatile memory}}.
\newblock {\emph{\JournalTitle{Scientific reports}}} \textbf{\bibinfo{volume}{6}}, \bibinfo{pages}{22616} (\bibinfo{year}{2016}).

\bibitem{zhu2023non}
\bibinfo{author}{Zhu, R.} \emph{et~al.}
\newblock \bibinfo{journal}{\bibinfo{title}{Non-volatile optoelectronic memory based on a photosensitive dielectric}}.
\newblock {\emph{\JournalTitle{Nature Communications}}} \textbf{\bibinfo{volume}{14}}, \bibinfo{pages}{5396} (\bibinfo{year}{2023}).

\bibitem{cheung2023ultra}
\bibinfo{author}{Cheung, S.} \emph{et~al.}
\newblock \bibinfo{journal}{\bibinfo{title}{Ultra-power-efficient electrically programmable photonic memory on a heterogeneous iii-v/si optical computing platform}}.
\newblock {\emph{\JournalTitle{Research Square preprint}}}  (\bibinfo{year}{2023}).

\bibitem{cheung2023non}
\bibinfo{author}{Cheung, S.} \emph{et~al.}
\newblock \bibinfo{title}{Non-volatile iii-v/si photonic charge-trap flash memory}.
\newblock In \emph{\bibinfo{booktitle}{2023 IEEE Photonics Conference (IPC)}}, \bibinfo{pages}{1--2} (\bibinfo{organization}{IEEE}, \bibinfo{year}{2023}).

\bibitem{tossoun2020memristor}
\bibinfo{author}{Tossoun, B.}, \bibinfo{author}{Sheng, X.}, \bibinfo{author}{Strachan, J.} \& \bibinfo{author}{Liang, D.}
\newblock \bibinfo{title}{The memristor laser}.
\newblock In \emph{\bibinfo{booktitle}{2020 IEEE International Electron Devices Meeting (IEDM)}}, \bibinfo{pages}{7--6} (\bibinfo{organization}{IEEE}, \bibinfo{year}{2020}).

\bibitem{cheung2022heterogeneous}
\bibinfo{author}{Cheung, S.} \emph{et~al.}
\newblock \bibinfo{title}{Heterogeneous iii-v/si non-volatile optical memory: A mach-zehnder memristor}.
\newblock In \emph{\bibinfo{booktitle}{CLEO: Science and Innovations}}, \bibinfo{pages}{STu5G--6} (\bibinfo{organization}{Optica Publishing Group}, \bibinfo{year}{2022}).

\bibitem{youngblood2023integrated}
\bibinfo{author}{Youngblood, N.}, \bibinfo{author}{R{\'\i}os~Ocampo, C.~A.}, \bibinfo{author}{Pernice, W.~H.} \& \bibinfo{author}{Bhaskaran, H.}
\newblock \bibinfo{journal}{\bibinfo{title}{Integrated optical memristors}}.
\newblock {\emph{\JournalTitle{Nature Photonics}}} \bibinfo{pages}{1--12} (\bibinfo{year}{2023}).

\bibitem{cheung2023nonGFP}
\bibinfo{author}{Cheung, S.} \emph{et~al.}
\newblock \bibinfo{title}{Non-volatile memristive iii-v/si photonics}.
\newblock In \emph{\bibinfo{booktitle}{2023 IEEE Silicon Photonics Conference (SiPhotonics)}}, \bibinfo{pages}{1--2} (\bibinfo{organization}{IEEE}, \bibinfo{year}{2023}).

\bibitem{hu2023inkjet}
\bibinfo{author}{Hu, H.} \emph{et~al.}
\newblock \bibinfo{journal}{\bibinfo{title}{Inkjet-printed tungsten oxide memristor displaying non-volatile memory and neuromorphic properties}}.
\newblock {\emph{\JournalTitle{Advanced Functional Materials}}} \bibinfo{pages}{2302290} (\bibinfo{year}{2023}).

\bibitem{stathis2006negative}
\bibinfo{author}{Stathis, J.~H.} \& \bibinfo{author}{Zafar, S.}
\newblock \bibinfo{journal}{\bibinfo{title}{The negative bias temperature instability in mos devices: A review}}.
\newblock {\emph{\JournalTitle{Microelectronics Reliability}}} \textbf{\bibinfo{volume}{46}}, \bibinfo{pages}{270--286} (\bibinfo{year}{2006}).

\bibitem{hehenberger2011advanced}
\bibinfo{author}{Hehenberger, P.~P.}
\newblock \emph{\bibinfo{title}{Advanced characterization of the bias temperature instability}}.
\newblock Ph.D. thesis, \bibinfo{school}{Technische Universit{\"a}t Wien} (\bibinfo{year}{2011}).

\bibitem{rzepa2014physical}
\bibinfo{author}{Rzepa, G.} \emph{et~al.}
\newblock \bibinfo{title}{Physical modeling of nbti: From individual defects to devices}.
\newblock In \emph{\bibinfo{booktitle}{2014 International Conference on Simulation of Semiconductor Processes and Devices (SISPAD)}}, \bibinfo{pages}{81--84} (\bibinfo{organization}{IEEE}, \bibinfo{year}{2014}).

\bibitem{entner2007modeling}
\bibinfo{author}{Entner, R.}
\newblock \emph{\bibinfo{title}{Modeling and simulation of negative bias temperature instability}}.
\newblock Ph.D. thesis, \bibinfo{school}{Technische Universit{\"a}t Wien} (\bibinfo{year}{2007}).

\bibitem{fussel1996defects}
\bibinfo{author}{F{\"u}ssel, W.}, \bibinfo{author}{Schmidt, M.}, \bibinfo{author}{Angermann, H.}, \bibinfo{author}{Mende, G.} \& \bibinfo{author}{Flietner, H.}
\newblock \bibinfo{journal}{\bibinfo{title}{Defects at the si/sio2 interface: their nature and behaviour in technological processes and stress}}.
\newblock {\emph{\JournalTitle{Nuclear Instruments and Methods in Physics Research Section A: Accelerators, Spectrometers, Detectors and Associated Equipment}}} \textbf{\bibinfo{volume}{377}}, \bibinfo{pages}{177--183} (\bibinfo{year}{1996}).

\bibitem{yuan2023mechanisms}
\bibinfo{author}{Yuan, Y.} \emph{et~al.}
\newblock \bibinfo{journal}{\bibinfo{title}{Mechanisms of enhanced sub-bandgap absorption in high-speed all-silicon avalanche photodiodes}}.
\newblock {\emph{\JournalTitle{Photonics Research}}} \textbf{\bibinfo{volume}{11}}, \bibinfo{pages}{337--346} (\bibinfo{year}{2023}).

\bibitem{silvaco2010}
\bibinfo{journal}{\bibinfo{title}{Simulating the hysteresis effects of si/sio2 interface traps}}.
\newblock {\emph{\JournalTitle{Silvaco Inc}}}  (\bibinfo{year}{2010}).

\bibitem{baba1986mechanism}
\bibinfo{author}{Baba, S.}, \bibinfo{author}{Kita, A.} \& \bibinfo{author}{Ueda, J.}
\newblock \bibinfo{title}{Mechanism of hot carrier induced degradation in mosfet's}.
\newblock In \emph{\bibinfo{booktitle}{1986 International Electron Devices Meeting}}, \bibinfo{pages}{734--737} (\bibinfo{organization}{IEEE}, \bibinfo{year}{1986}).

\bibitem{hofmann1985hot}
\bibinfo{author}{Hofmann, K.~R.}, \bibinfo{author}{Werner, C.}, \bibinfo{author}{Weber, W.} \& \bibinfo{author}{Dorda, G.}
\newblock \bibinfo{journal}{\bibinfo{title}{Hot-electron and hole-emission effects in short n-channel mosfet's}}.
\newblock {\emph{\JournalTitle{IEEE Transactions on Electron Devices}}} \textbf{\bibinfo{volume}{32}}, \bibinfo{pages}{691--699} (\bibinfo{year}{1985}).

\bibitem{tan2000avalanche}
\bibinfo{author}{Tan, C.} \emph{et~al.}
\newblock \bibinfo{journal}{\bibinfo{title}{Avalanche noise measurement in thin si p+-in+ diodes}}.
\newblock {\emph{\JournalTitle{Applied Physics Letters}}} \textbf{\bibinfo{volume}{76}}, \bibinfo{pages}{3926--3928} (\bibinfo{year}{2000}).

\bibitem{kang2009monolithic}
\bibinfo{author}{Kang, Y.} \emph{et~al.}
\newblock \bibinfo{journal}{\bibinfo{title}{Monolithic germanium/silicon avalanche photodiodes with 340 ghz gain--bandwidth product}}.
\newblock {\emph{\JournalTitle{Nature photonics}}} \textbf{\bibinfo{volume}{3}}, \bibinfo{pages}{59--63} (\bibinfo{year}{2009}).

\end{thebibliography}



\section*{Author contributions statement}
Y.Y. designed the devices, taped out the chips, performed the measurements, built the model, and wrote the manuscript. Y.P., S.C., and W.V.S. participated in the measurements, Z.H., D.L., M.F., and R.G.B supervised the study and gave important technical advice. All authors reviewed the manuscript. 

\section*{Competing interests statement}
The authors declare no competing interests.

\end{document}


\flushbottom
\maketitle

\thispagestyle{empty}

\renewcommand{\thefigure}{S\arabic{figure}}
\setcounter{figure}{0}

\section*{I. Non-radiative multi-phonon process}

\begin{figure}[ht]
\centering
\includegraphics[width=0.75\linewidth]{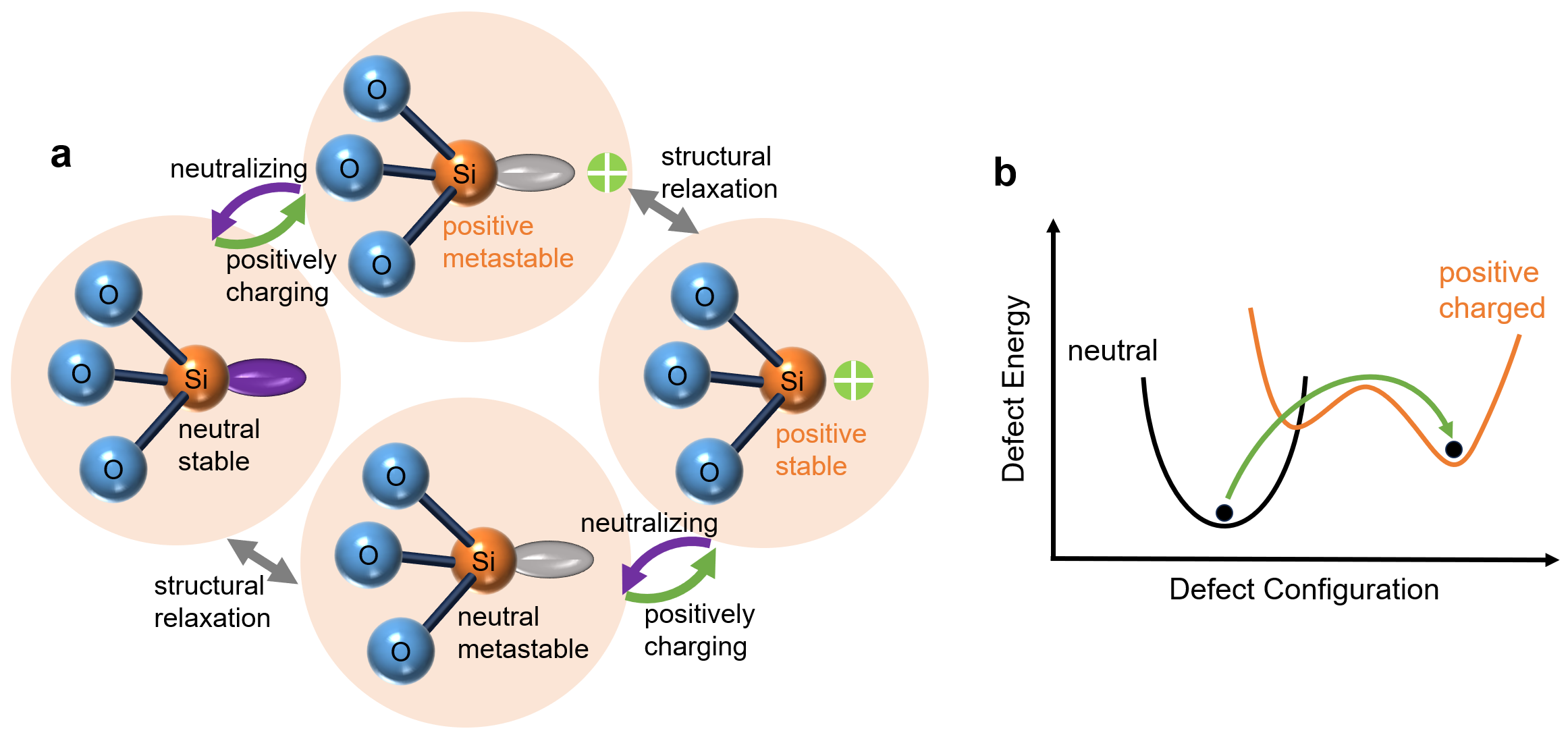}
\caption{\textbf{Non-radiative multi-phonon (NMP) process.} \textbf{a}, A four-state (neutral stable, positive metastable, positive stable, and natural metastable states) model with charge exchange and structural relaxation processes. \textbf{b}, Schematic illustration of energy diagram for an NMP process.}
\label{fig.s1}
\end{figure}

The resonant wavelength of the Si microring resonator (MRR) memory undergoes a red shift after programming. This shift wavelength, donated as $\Delta\lambda$, increases with higher temperatures. It suggests that hole trapping is a temperature-dependent process, which is not some form of elastic tunneling, as tunneling is temperature-independent. This hole trapping is explicable through a non-radiative multi-phonon (NMP) process. The NMP model comprises a four-state system, consisting of two stable states and two metastable states, as depicted in Fig. \ref{fig.s1}a. Transitions between the neutral stable state and the positive stable state need to pass through a metastable state. A stable state requires extra structural relaxation in addition to neutralization or charging \cite{hehenberger2011advanced,schanovsky2012multi}. The energy diagram of the neutral (black) and the positive charged (orange) defect is illustrated in Fig. \ref{fig.s1}b. During an NMP process, the defect needs to thermally overcome the energy barrier to capture or emit a charge carrier. Therefore, the non-volatile shift of the Si optical memory improves with higher temperatures.

\section*{II. Hero device}

\begin{figure}[ht]
\centering
\includegraphics[width=0.7\linewidth]{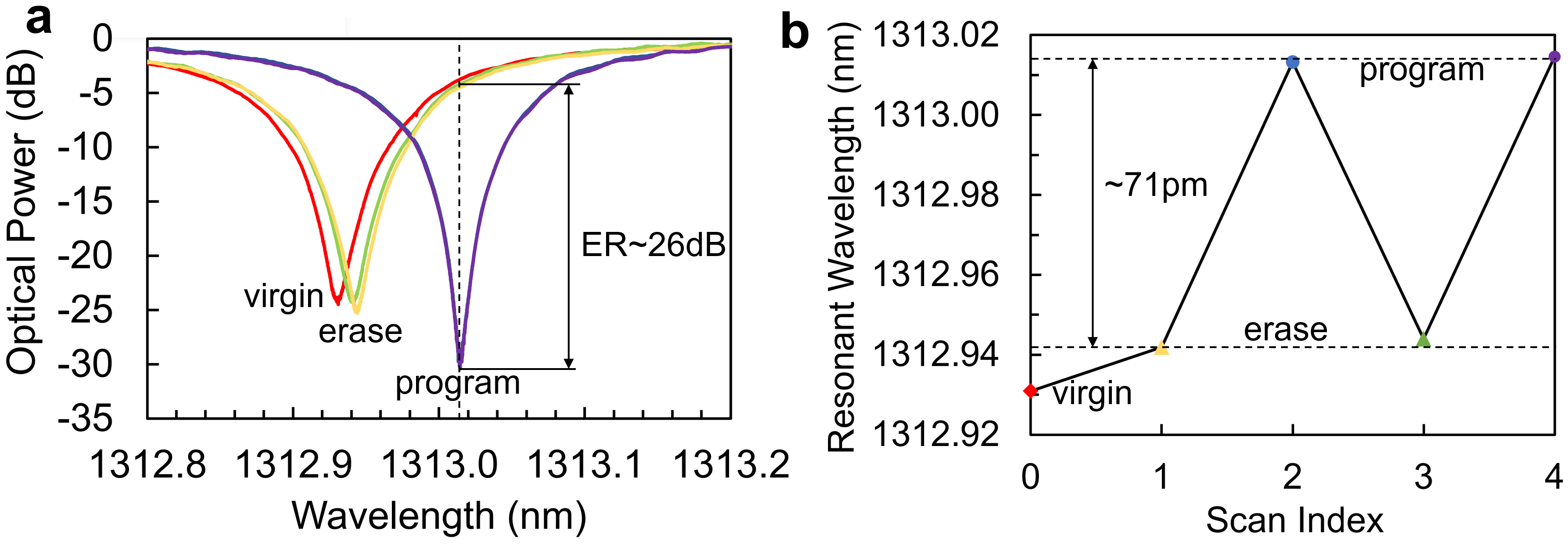}
\caption{\textbf{Characteristics of the hero Si optical memory device.} \textbf{a}, Measured optical spectrum of the virgin, erase, and program states of the hero device. \textbf{b}, Corresponding resonant wavelength of the hero device.}
\label{fig.s2}
\end{figure}

The two Si optical memories represented in the main text exhibit a similar non-volatile resonant wavelength shift $\Delta\lambda$ of $\sim$ 42 pm. However, this $\Delta\lambda$ value is the median value of the memories on the wafer. Since the $\Delta\lambda$ is determined by the trapping charge, some devices with higher trap density could result in greater variations. Figure \ref{fig.s2} illustrates the characteristics of the hero device on the same wafer. Similar to Fig. 5, the hero device was also measured in virgin, erase, and program states. The optical spectrum of these states is depicted in Fig. \ref{fig.s2}a, where the virgin state exhibits the shortest resonant wavelength, and the program-erase cycles are repeatable. At the resonant wavelength of the program state, this hero optical memory exhibits a remarkably high extinction ratio (ER) of $\sim$ 26 dB. The corresponding resonant wavelengths are plotted in Fig. \ref{fig.s2}b, the repeatable wavelength shift between program and erase states is about 71 pm. This shift can be further enhanced by carefully optimizing the device design and fabrication process.

\section*{III. Optical memory fabricated at Leti}

\begin{figure}[ht]
\centering
\includegraphics[width=1\linewidth]{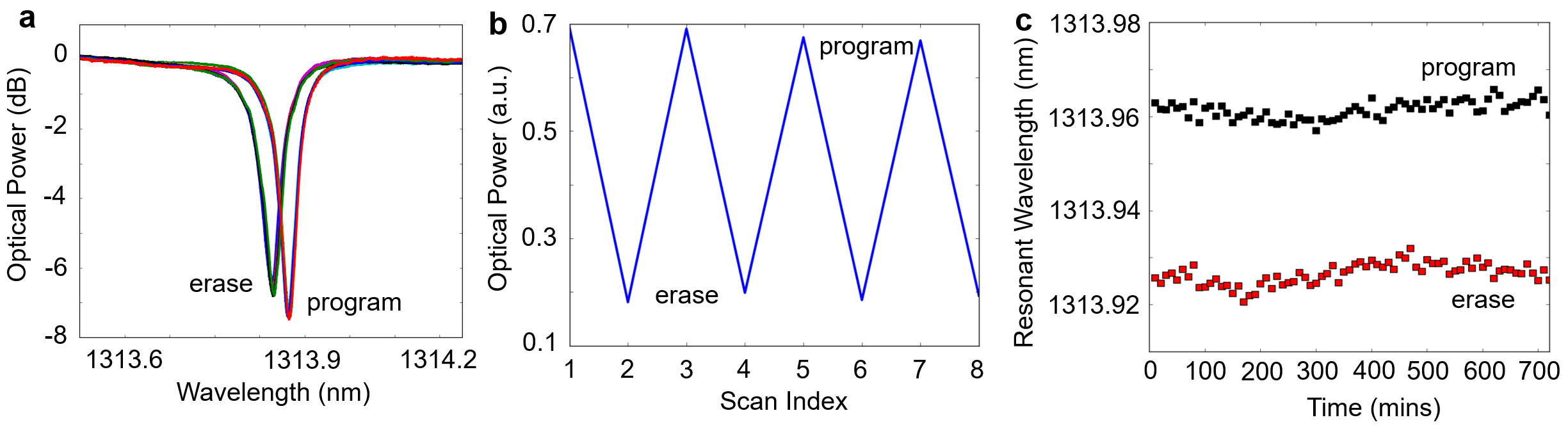}
\caption{\textbf{Characteristics of the Si optical memory fabricated at Leti.} \textbf{a}, Measured optical spectrum of the Leti optical memory at 75 $\tccentigrade$. \textbf{b}, Corresponding optical power switches in a linear scale. \textbf{c}, Retention measurements of the non-volatile resonant wavelengths of the program and erase states within a 12-hour period.}
\label{fig.s3}
\end{figure}

The proposed Si optical memory is a generic device that is not restricted by the specifications of a particular Si photonics foundry. The aforementioned results are based on the Si MRR devices fabricated at Advanced Micro Foundry (AMF), but we have also successfully reproduced the non-volatile phenomena in a Si MRR device fabricated at CEA-Leti. Due to the independent processes of these different foundries, the consistent results suggest that the performance of the proposed Si optical memory is not coincidental. The memory characteristic of the Leti device is depicted in Fig. \ref{fig.s3}, and all results were measured at 75 $\tccentigrade$ as well. The optical spectrum of the program and erase states are shown in Fig. \ref{fig.s3}a, this device exhibits repeatable switches between two states. Since the doping concentrations of this device are lower than that of AMF MRR, it boasts a higher quality factor (Q). It thus enables a substantial optical power switch, as shown in Fig. \ref{fig.s3}b, which offers about 50\% power switch in a linear scale. Moreover, the resonant wavelengths of this Leti optical memory were measured every 10 minutes over 12 hours for both program and erase states, demonstrating stable retention capability.

\bibliography{supplementary}